\documentclass{article}
\pdfoutput=1
\usepackage{spconf,amsmath,graphicx}
\usepackage{fancyhdr}
\usepackage{graphicx} 
\usepackage{amssymb,amsfonts}
\usepackage{IEEEtrantools}
\usepackage{import}  %
\usepackage{siunitx}    %
\usepackage[american]{circuitikz} %
\usepackage{tikz}
\usetikzlibrary{fit,matrix,chains,positioning,decorations.pathreplacing,arrows,
external}
\usepackage{standalone}  %
\usepackage{pgfplots} 
\pgfplotsset{table/search path={inc},compat=1.16}
\usepackage[section]{placeins} %
\usepackage{balance}

\newcommand{\bma}{\begin{bmatrix}}
\newcommand{\ema}{\end{bmatrix}}
\newcommand{\T}{{\mathsf{T}}} %
\newcommand{\Reals}{\mathbb{R}}      %
\newcommand{\dd}{\mathop{}\!\mathrm{d}} %
\newcommand{\eqdef}{\triangleq} %

\newcommand{\argmax}{\operatorname*{argmax}}
\makeatletter
\DeclareFontFamily{U}{MnSymbolA}{}
\DeclareSymbolFont{MnSyA}{U}{MnSymbolA}{m}{n}
\DeclareFontShape{U}{MnSymbolA}{m}{n}{
<-6> MnSymbolA5
<6-7> MnSymbolA6
<7-8> MnSymbolA7
<8-9> MnSymbolA8
<9-10> MnSymbolA9
<10-12> MnSymbolA10
<12-> MnSymbolA12}{}
\DeclareMathSymbol{\smallrightarrow}{\mathrel}{MnSyA}{0}
\DeclareMathSymbol{\smallleftarrow}{\mathrel}{MnSyA}{2}
\DeclareMathSymbol{\smallleftrightarrow}{\mathrel}{MnSyA}{16}
\newcommand{\smallrightarrowfill@}{\arrowfill@\relbar\relbar\smallrightarrow}
\newcommand{\smallleftarrowfill@}{\arrowfill@\smallleftarrow\relbar\relbar}
\newcommand{\smallleftrightarrowfill@}
{\arrowfill@\smallleftarrow\relbar\smallrightarrow}
\renewcommand{\overrightarrow}{\mathpalette{\overarrow@\smallrightarrowfill@}}
\renewcommand{\overleftarrow}{\mathpalette{\overarrow@\smallleftarrowfill@}}
\renewcommand{\overleftrightarrow}
{\mathpalette{\overarrow@\smallleftrightarrowfill@}}
\makeatother

\DeclareMathOperator{\E}{\textnormal{\ensuremath{\mathbb{E}}}}
\newcommand{\EE}[1]{\E\!\left[{#1}\right]}

\newcommand{\cond}{\hspace{0.02em}|\hspace{0.08em}}
\newcommand{\Normal}[1]{\mathcal{N}\!\left({#1}\right)} %

\title{BINARY CONTROL AND DIGITAL-TO-ANALOG CONVERSION USING 
COMPOSITE NUV PRIORS AND ITERATIVE GAUSSIAN MESSAGE PASSING}
\name{Raphael Keusch, Hampus Malmberg, and Hans-Andrea Loeliger}
\address{ETH Zurich, Dept. of Information Technology \& Electrical Engineering \\
\{keusch, malmberg, loeliger\}@isi.ee.ethz.ch}

\fancypagestyle{plain}{%
   \fancyhf{} 
   \fancyfoot[C]{}
   
}

\begin{document}

\maketitle

\begin{abstract}
The paper proposes a new method to determine a binary 
control signal for an analog linear system such that the state, or some output, 
of the system follows a given target trajectory.
The method can also be used for digital-to-analog conversion.

The heart of the proposed method is a new binary-enforcing 
NUV prior (normal with unknown variance). 
The resulting computations, for each planning period, 
amount to iterating forward-backward Gaussian message passing recursions
(similar to Kalman smoothing),
with a complexity (per iteration) that is linear in the planning horizon.
In consequence, 
the proposed method is not limited to a short planning horizon. 
\end{abstract}

\section{Introduction}
\label{sec:introduction}

Consider the classical control problem 
of steering an analog physical linear system along some 
desired trajectory, or to make the system produce some desired analog output.
In this paper, we are interested in the special case 
where the control input is binary (i.e., restricted to two%
\footnote{
The method of this paper can be extended to control signals 
with $M>2$ levels, as will be described elsewhere.} 
levels), 
which makes the problem much harder.
This binary-input control problem includes, in particular, 
a certain type of analog-to-digital converter 
where the binary output of some digital processor directly drives 
a continuous-time analog linear filter---preferably an inexpensive one---%
which produces the desired analog waveform.

It is tempting to ask for an optimal binary control signal,
i.e., a control signal that produces the best approximation 
of the desired analog trajectory 
(e.g., for a quadratic cost function). 
However, determining such an optimal control signal 
is a hard combinatorial optimization problem,
with a computational complexity growing exponentially 
with the planning horizon \cite{land_automatic_1960,wolsey_integer_1999}. 
In consequence, insisting on an optimal control signal 
limits us to a short planning horizon, which is a very severe restriction.
This problem is well known in model predictive control (MPC)
\cite{aguilera_stability_2011,geyer_multistep_2014}.
Techniques such as sphere decoding do help 
\cite{dorfling_long-horizon_2019}, 
but the fundamental problem remains.

Clearly, the binary-input control problem is a nonconvex optimization problem.
A general approach to nonconvex optimization 
is to resort to some convex relaxation, 
and to project the solution back to the permissible set \cite{sparrer_adapting_2014}.
Other general approaches 
include heuristic methods such as random-restart 
hill-climbing~\cite{russel_artificial_2013} and
simulated annealing~\cite{pincus_letter_1970}. 

The heart of the method proposed in this paper is a new binary-enforcing 
NUV prior (normal with unknown variance).
NUV priors are a central idea of sparse Bayesian learning 
\cite{tipping_sparse_2001, tipping_fast_2003,wipf_sparse_2004,wipf_new_2008},
and closely related to variational representations of Lp norms~\cite{loeliger_factor_2018,bach_optimization_2012}.
Such priors have been used mainly for sparsity; 
in particular, no binary-enforcing NUV prior seems to have been proposed in the prior literature.
(An interesting non-NUV binary-enforcing prior has been proposed in \cite{dai_sparse_2019}.)

A main advantage of NUV priors in general is their computational compatibility 
with linear Gaussian models, cf.~\cite{loeliger_sparsity_2016}.
In this paper, the computations (for each planning period)
amount to iterating forward-backward Gaussian message passing recursions
similar to Kalman smoothing,
with a complexity (per iteration) that is linear in the planning horizon.
In consequence, the proposed method can effectively 
handle long planning horizons, which can far outweigh its suboptimality.

\section{The Binary-enforcing NUV Prior}
\label{sec:PriorModel}

Let $\Normal{x; \mu, \sigma^2}$ 
denote the normal probability density function in $x$ 
with mean $\mu\in\Reals$ and variance $\sigma^2$.
Let 
\begin{IEEEeqnarray}{rCl} 
\label{eqn:TwoLvlPrior}
\rho(x, \theta) \eqdef \Normal{x; a, \sigma_1^2} \Normal{x; b, \sigma_2^2},
\end{IEEEeqnarray}
where $\theta \eqdef (\sigma_1^2, \sigma_2^2)$
is a shorthand for the two unknown variances in (\ref{eqn:TwoLvlPrior}).
For fixed $\theta$,
$\rho(x, \theta)$ is a normal probability density in $x$ (up to a scale factor), 
which is essential for the algorithms in Section~\ref{sec:Algorithms}.

\begin{figure}[t]
\setlength{\unitlength}{0.9mm}
\newcommand{\cent}[1]{\makebox(0,0){#1}}
\newcommand{\pos}[2]{\makebox(0,0)[#1]{#2}}
\newcommand{\knownBox}{\cent{\rule{1.75\unitlength}{1.75\unitlength}}}
\newcommand{\calN}{\ensuremath{\mathcal N}}
\centering
\begin{picture}(75,45)(0,0)

\put(0,30){\vector(1,0){10}}  \put(0,31){\pos{bl}{$\sigma_1$}}
\put(10,27.5){\framebox(5,5){$\times$}}
 \put(12.5,37.5){\vector(0,-1){5}}
 \put(10,37.5){\framebox(5,5){\calN}}
\put(15,30){\vector(1,0){7.5}}
\put(22.5,27.5){\framebox(5,5){$+$}}
 \put(25,40){\vector(0,-1){7.5}}
 \put(25,40){\knownBox}  \put(27,40){\pos{cl}{$a$}}
\put(27.5,30){\line(1,0){7.5}}
 \put(27.5,30){\vector(1,0){5}}
\put(35,30){\line(0,-1){5}}
\put(0,15){\vector(1,0){10}}  \put(0,16){\pos{bl}{$\sigma_2$}}
\put(10,12.5){\framebox(5,5){$\times$}}
 \put(12.5,7.5){\vector(0,1){5}}
 \put(10,2.5){\framebox(5,5){\calN}}
\put(15,15){\vector(1,0){7.5}}
\put(22.5,12.5){\framebox(5,5){$+$}}
 \put(25,5){\vector(0,1){7.5}}
 \put(25,5){\knownBox}   \put(27,5){\pos{cl}{$b$}}
\put(27.5,15){\line(1,0){7.5}}
 \put(27.5,15){\vector(1,0){5}}
\put(35,15){\line(0,1){5}}
\put(32.5,20){\framebox(5,5){$=$}}
\put(5,0){\dashbox(35,45){}}   \put(41,0){\pos{bl}{$\rho(x,\theta)$}}

\put(37.5,22.5){\vector(1,0){17.5}}  \put(45,23.6){\pos{cb}{$X$}}

 \put(55,30){\framebox(5,5){}}  \put(57.5,36){\pos{cb}{$\Normal{0,s^2}$}}
 \put(57.5,30){\vector(0,-1){5}}
\put(55,20){\framebox(5,5){$+$}}
\put(60,22.5){\line(1,0){7}}
 \put(60,22.5){\vector(1,0){3.5}}
\put(67,22.5){\knownBox}     \put(69,22.5){\pos{cl}{$\mu$}}
\put(49,15){\dashbox(23.5,30){}}  \put(61,13.5){\pos{ct}{$p(\breve y \cond x)$}}
\end{picture}
\caption{\label{fig:BinaryPriorWithGaussianLikelihood}%
Factor graph of (\ref{eqn:BinaryPriorWithGaussianLikelihood})
for fixed $\breve y$.
The boxes labeled ``\calN'' 
represent normal probability density functions $\Normal{0,1}$.
}
\end{figure}
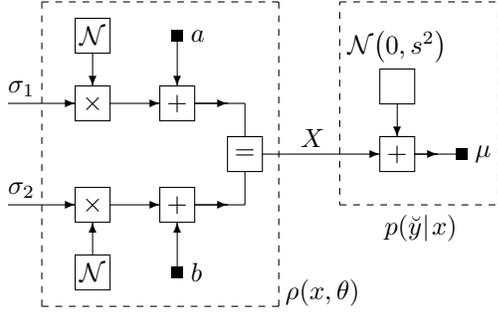

\begin{figure}
\centering
\includegraphics{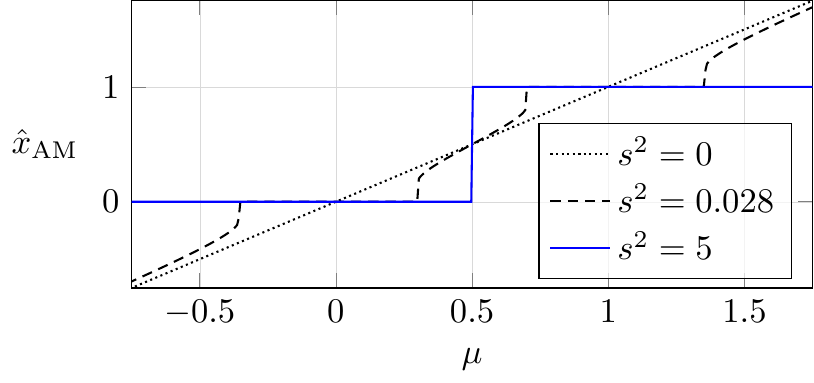}
\vspace{-3ex}
\caption{\label{fig:hatXJointMAP}%
The estimate of Section~\ref{sec:Prior:JointMAP},
for $a=0$ and $b=1$, as a function of $\mu$.}
\vspace{\dblfloatsep}

\includegraphics{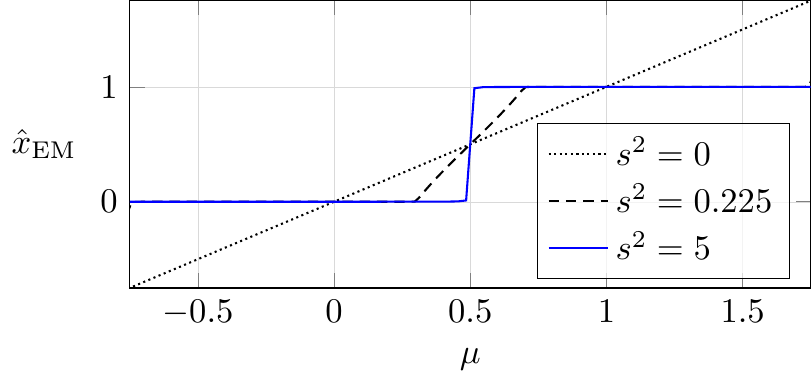}
\vspace{-3ex}
\caption{\label{fig:hatXTypeII}%
The estimate of Section~\ref{sec:Prior:VarMAP},  
for $a=0$ and $b=1$, as a function of $\mu$.}
\end{figure}

The heart of the proposed method is
the observation that~(\ref{eqn:TwoLvlPrior})
can be used as a (improper) joint prior for $x$ and $\theta$ 
that strongly encourages $x$ to lie in $\{ a, b \}$. 
To see this, assume $\rho(x, \theta)$ is used 
in some model with (fixed) observation(s) $\breve y$
and likelihood function $p(\breve y \cond x)$.
Two different ways to estimate the variances~$\theta$
are considered in Sections \ref{sec:Prior:JointMAP} and~\ref{sec:Prior:VarMAP}.
For the numerical examples 
in Figs.\ \ref{fig:hatXJointMAP} and~\ref{fig:hatXTypeII},
we will assume that $p(\breve y \cond x)$  
is Gaussian (in $x$) with mean $\mu$ and variance $s^2$
depending on $\breve y$,
i.e.,
\begin{equation} \label{eqn:prior:GaussianLikelihood}
p(\breve y \cond x) = \Normal{x; \mu, s^2}.
\end{equation}
A factor graph \cite{loeliger_introduction_2004}
of the resulting statistical system model
\begin{equation} \label{eqn:BinaryPriorWithGaussianLikelihood}
p(\breve y \cond x) \rho(x, \theta)
= \Normal{x; \mu, s^2} \Normal{x; a, \sigma_1^2} \Normal{x; b, \sigma_2^2}
\end{equation}
is shown in Fig.~\ref{fig:BinaryPriorWithGaussianLikelihood}.

\subsection{Joint MAP Estimation}
\label{sec:Prior:JointMAP}

Assume that $x$ and $\theta$ are determined by joint MAP estimation. 
The resulting estimate $\hat x$ of $x$ is
\begin{IEEEeqnarray}{rCl}
\hat x 
& = & \argmax_{x} \max_{\theta} p(\breve y \cond x) \rho(x, \theta) \\
& = & \argmax_{x} 
      p(\breve y \cond x) \max_{\theta} \rho(x, \theta),
      \label{eqn:hatXJointMAP2}
      \IEEEeqnarraynumspace
\end{IEEEeqnarray}
with an effective (improper) prior
\begin{IEEEeqnarray}{rCl}
\max_{\theta} \rho(x, \theta)
& = & 
\max_{\sigma_1^2} \Normal{x; a, \sigma_1^2}
\max_{\sigma_2^2} \Normal{x; b, \sigma_2^2} \IEEEeqnarraynumspace\\
& \propto & \frac{1}{|x-a|\cdot |x-b|}
\label{eqn:jointMAPprior}
\end{IEEEeqnarray}
It is obvious that this effective prior 
has a strong preference for $x$ to lie in $\{ a, b\}$.

Such estimates may conveniently be computed by alternating maximization (AM)
over $x$ and $\theta$, cf.\ Section~\ref{sec:Algo:AM}.
However, AM may converge to a local maximum 
rather than the global maximum (\ref{eqn:hatXJointMAP2}),
in which case the estimate $\hat x_\text{AM}$ returned by AM
may lie outside $\{ a, b\}$.

For $p(\breve y \cond x)$ as in (\ref{eqn:prior:GaussianLikelihood}),
the estimate $\hat x_\text{AM}$ is plotted in Fig.~\ref{fig:hatXJointMAP}.
In this setting, we observe (and it can be proved) 
that, for every fixed $\mu$ and sufficiently large $s^2$,
there is no local maximum and $\hat x_\text{AM}$ lies in $\{ a, b \}$.

\subsection{Type-II Estimation\protect\footnote{in the sense of \cite{tipping_sparse_2001, wipf_sparse_2004}}}
\label{sec:Prior:VarMAP}

In this case, 
we first determine the MAP estimate $\hat\theta$ of $\theta$, i.e.,
\begin{equation} \label{eqn:hatThetaMAP}
\hat\theta = \argmax_{\theta} \int_{-\infty}^\infty p(\breve y \cond x) \rho(x, \theta)\, dx,
\end{equation}
and then we estimate $x$ as
\begin{equation} \label{eqn:hatXTypeII}
\hat x = \argmax_{x} p(\breve y \cond x) \rho(x, \hat\theta).
\end{equation}

Such estimates may conveniently be computed by expectation maximization (EM) 
(cf.\ Section~\ref{sec:Algo:EM}), which, however, may converge to a local maximum. 
Moreover, even the global maximum (\ref{eqn:hatXTypeII}) may lie outside $\{ a, b \}$.
Nonetheless, 
for $p(\breve y \cond x)$ as in (\ref{eqn:prior:GaussianLikelihood}),
we observe (and it can be proved) that, 
for every fixed $\mu \neq (a+b)/2$ and sufficiently large $s^2$,
$\hat x_\text{EM}$ lies in $\{ a, b \}$,
as illustrated in  Fig.~\ref{fig:hatXTypeII}.

\section{System Model and Algorithms}
\label{sec:SystemModelAlgo}

\subsection{Problem Statement}
\label{sec:StateSpaceProblem}

Consider a linear system with scalar input $u_k\in\Reals$ and state $x_k\in\Reals^N$,
which evolves according to 
\begin{equation} \label{eqn:StateSpaceEvolution}
x_k  = A x_{k-1} + B u_k,
\end{equation}
where $k\in \{ 1, 2, \ldots, K\} $
is the time index (with finite horizon $K$),
and where 
both $A\in\Reals^{N\times N}$ and $B\in\Reals^{N\times 1}$ are assumed to be known.
Our goal is to determine a two-level input signal 
$u_1,\, \ldots, u_K \in \{a, b\}$ 
such that some output (or feature) 
\begin{equation} \label{eqn:StateSpaceOutput}
y_k = C x_k \in \Reals^L
\end{equation}
(with known $C\in\Reals^{L\times N}$) 
follows a given trajectory $\breve y_1$, \ldots, $\breve y_K \in\Reals^L$,
i.e., we wish
\begin{equation} \label{eqn:trajectoryCost}
\sum_{k=1}^K \|y_k - \breve y_k\|^2
\end{equation}
to be as small as possible.
For ease of exposition, we will assume that the initial state $x_0$ is known.

Note that this off\-line control problem may be viewed 
as a single episode of an online control problem, with planning horizon $K$.
Note also that we are primarily interested in $K\gg 1$,
which precludes exhaustive tree search algorithms.

\subsection{The Statistical Model}
\label{sec:StatModel}

In order to solve the problem stated in Section~\ref{sec:StateSpaceProblem},
we turn it into a statistical estimation problem 
with an (improper) i.i.d.\ prior%
\footnote{In (\ref{eqn:StatModelPrior}), $\rho$ is used for two different functions (with different arguments).} 
\begin{equation} \label{eqn:StatModelPrior}
  \rho(u, \theta) \eqdef \prod_{k=1}^K \rho(u_k, \theta_k),
\end{equation}
where $u\eqdef (u_1,\, \ldots, u_K)$,
$\theta \eqdef (\theta_1,\, \ldots, \theta_K)$,
and 
\begin{equation} \label{eqn:BinSequencePrior}
\rho(u_k, \theta_k) \eqdef \Normal{u_k; a, \sigma_{1,k}^2} \Normal{u_k; b, \sigma_{2,k}^2}
\end{equation}
with $\theta_k = (\sigma_{1,k}^2, \sigma_{2,k}^2)$
as in (\ref{eqn:TwoLvlPrior}).
Accordingly, we replace (\ref{eqn:trajectoryCost}) by the likelihood function
\begin{equation}
p(\breve y \cond u)
= \prod_{k=1}^K \frac{1}{(2\pi)^{L/2}s^L} 
                \exp\left( \frac{- \| y_k - \breve y_k \|^2}{2s^2} \right),
\end{equation}
where $\breve y = (\breve y_1,\, \ldots, \breve y_K)$
and where $s>0$ is a free parameter.
The complete statistical model is then given by 
\begin{equation}
p(\breve y, u, \theta) \eqdef p(\breve y \cond u) \rho(u, \theta)
\end{equation}
together with (\ref{eqn:StateSpaceEvolution}) and~(\ref{eqn:StateSpaceOutput}).

\subsection{Algorithms}
\label{sec:Algorithms}

Both joint MAP estimation of $u$ and $\theta$ (as in Section~\ref{sec:Prior:JointMAP})
and type-II estimation of $u$ and $\theta$ (as in Section~\ref{sec:Prior:VarMAP}) 
can be implemented as special cases 
(with different versions of Step~2)
of the following algorithm.

\subsubsection{Iterative Kalman Input Estimation (IKIE)}
\label{sec:IKIE}

The algorithm estimates $\theta$ and $u$
by alternating the following two steps for $i=1, 2, 3,\, \ldots\, $:
\begin{enumerate}
\item
  For fixed $\theta = \theta^{(i-1)}$,
  compute the posterior means $\hat u_k^{(i)}$ of $u_k$ (for $k=1,\, \ldots, K$) and,
  if necessary, the posterior variances $V_{U_k}^{(i)}$ of $u_k$,
  with respect to the probability distribution 
  $p(u \cond \breve y, \theta)$.
\item
  From these means and variances, determine new NUV parameters $\theta^{(i)}$.
\end{enumerate}
Note that Step~1 operates with a standard linear Gaussian model.
In consequence, the required means and variances can be computed 
by standard Kalman-type recursions or, equivalently, by forward-backward Gaussian 
message passing, with a complexity that is linear in $K$.

A preferred such algorithm is MBF message passing as in \cite[Section~V]{loeliger_sparsity_2016},
which amounts to Modified Bryson–Frazier smoothing~\cite{bierman_factorization_1977}
augmented with input signal estimation. 
This algorithm requires no matrix inversion%
\footnote{This is obvious for $L=1$. For $L>1$, a little adaptation is required.}
and is numerically quite stable.

\subsubsection{Determining $\theta$ and $u$ by Joint MAP Estimation}
\label{sec:Algo:AM}

In this case, we wish to compute the estimate
\begin{IEEEeqnarray}{rCl}
\hat u 
 & = &  \argmax_{u} \max_\theta p(\breve y \cond u) \rho(u, \theta). 
         \IEEEeqnarraynumspace\label{eqn:SequenceJointMAP}
\end{IEEEeqnarray}
The double maximization (over $u$ and $\theta$) is naturally implemented 
by alternating maximization, which can be implemented by the IKIE algorithm,
with Step~2 given by
\begin{equation} \label{eqn:AMUpdate}
\big( \sigma_{1,k}^2 \big)^{(i)} = \big( \hat u_{k}^{(i)}\, - a \big)^2
\text{~~and~~}
\big( \sigma_{2,k}^2 \big)^{(i)} = \big( \hat u_{k}^{(i)}\, - b \big)^2.
\end{equation}

\subsubsection{Type-II Estimation}
\label{sec:Algo:EM}

In this case, we wish to compute the estimate 
\begin{equation} \label{eqn:ThetaTypeII}
  \hat \theta = \argmax_{\theta} \int p(\breve y \cond u)\rho(u,\theta) \dd u,
\end{equation}
which can be carried out by expectation maximization~\cite{stoica_cyclic_2004}
with hidden variable(s) $u$. 
The update step for $\theta$ is 
\begin{IEEEeqnarray}{rCl} \label{eqn:ThetaUpdateEM}
  \theta^{(i)} &=& \argmax_{\theta} \EE{\log p(\breve y \cond U) \rho(U, \theta)}\\
  &=& \argmax_{\theta} \EE{\log \rho(U, \theta)},
\end{IEEEeqnarray}
where the expectation is with respect to $p(u \cond \breve y, \theta^{(i-1)})$.
The update (\ref{eqn:ThetaUpdateEM}) turns out to be computable by the IKIE algorithm
with Step~2 given by
\begin{IEEEeqnarray}{rCl}
\label{eqn:EMUpdateFinalDAC}
\big( \sigma_{1,k}^2 \big)^{(i)} 
& = &  V_{U_{k}}^{(i)} + \big(\hat u_{k}^{(i)} - a \big)^2 \text{~~and}
      \label{eqn:EMUpdateA}\\
\big( \sigma_{2,k}^2 \big)^{(i)} 
& = & V_{U_{k}}^{(i)} + \big(\hat u_{k}^{(i)} -b \big)^2 \!.
      \label{eqn:EMUpdateB} 
\end{IEEEeqnarray}

\subsubsection{Remarks}

\begin{enumerate}
\item
The algorithms of this section 
normally converge to a local (not the global) maximum 
of (\ref{eqn:SequenceJointMAP}) or (\ref{eqn:ThetaTypeII}).
\item
The parameter $s$ controls the error (\ref{eqn:trajectoryCost}).
If $s$ is chosen too small, the algorithm may return a nonbinary $u$. 

\item
Type-II estimation empirically works better than joint MAP estimation 
(in agreement with~\cite{giri_type_2016}). 
The numerical results in Figs.\ \ref{fig:DACWaveforms} and~\ref{fig:FlappyBirdTrajectory}
are obtained with type-II estimation.
\end{enumerate}

\section{Examples}
\label{sec:Examples}

\subsection{Digital-to-Analog Conversion}
\label{sec:ExDAC}

\begin{figure}[!t]
\begin{center}
\vspace{0.5ex}
\includegraphics{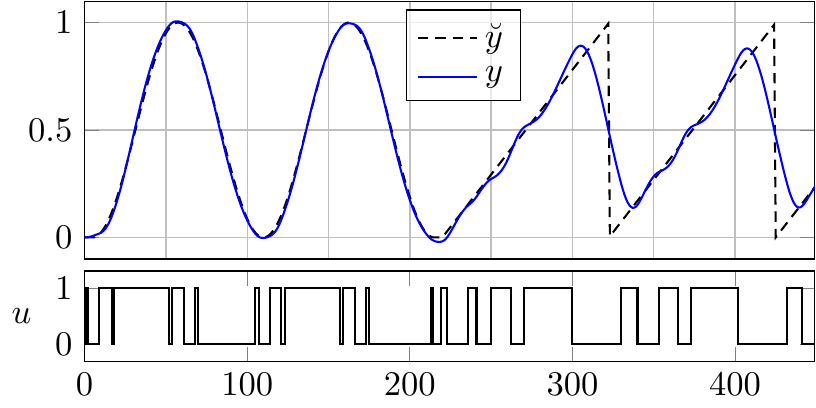}
\vspace{-1ex}
\caption{\label{fig:DACWaveforms}%
Digital-to-analog conversion as in Section~\ref{sec:ExDAC} 
with target waveform $\breve y$ (dashed), digital control signal $u$ (bottom), 
and filter output signal $y$ (solid blue).}
\end{center} 
\end{figure}

One method for digital-to-analog conversion 
is to feed a continuous-time analog linear filter
directly with a binary output signal $u$ 
of a digital processor. 
This method requires an algorithm to 
compute a suitable binary signal $u$ 
such that the analog filter output approximates 
the desired analog signal $\breve y$.
A standard approach is to compute $u$ by a delta-sigma modulator~\cite{boser_design_1988},
which requires the analog filter to approximate an ideal low-pass filter.
By contrast, the method of this paper works also with simpler (i.e., less expensive) 
analog filters.

For the following numerical example, the analog filter is a
3rd-order low-pass,   %
resulting in the discrete-time state space model 
\begin{IEEEeqnarray}{rCl}
    A &=& \bma
            0.7967&  -6.3978& -94.2123\\
0.0027 &  0.9902 & -0.1467\\
0      &  0.0030 &  0.9999
            \ema,
\end{IEEEeqnarray}
$B = \bma 0.0027 & 0 & 0  \ema^{\T}\!$, and 
$C = \bma 0 & 0 & 35037.9 \ema$.
The numerical results in Fig.~\ref{fig:DACWaveforms}
are obtained with $a=0$, $b=1$, $K=450$ and $s^2 = 0.045$.

\subsection{Trajectory Planning with Sparse Checkpoints}
\label{sec:ExTrajPlanning}

\begin{figure}
\begin{center}
\includegraphics{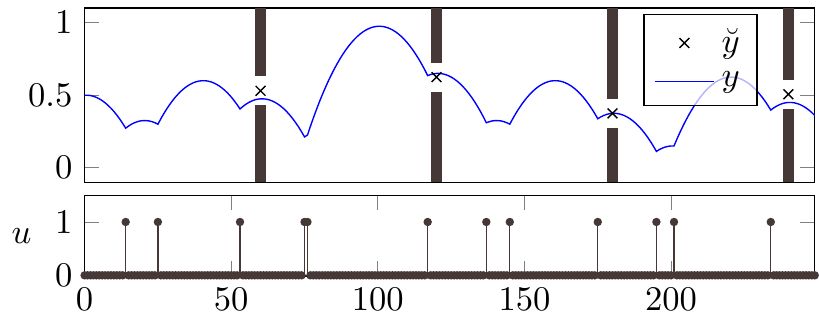}
\vspace{-0.5ex}
\caption{\label{fig:FlappyBirdTrajectory}%
Flappy bird control
with check points $\breve y$, binary control signal $u$ (bottom), 
and resulting trajectory $y$ (solid blue).}
\end{center} 
\end{figure}

The following control problem is a version of the \emph{flappy bird} computer game~\cite{wiki_flappy_bird}.
Consider an analog physical system consisting 
of a point mass $m$ moving forward (left to right in Fig.~\ref{fig:FlappyBirdTrajectory}) 
with constant horizontal velocity
and ``falling'' vertically with constant acceleration $g$. 
The $\{0,1\}$-valued control signal $u$ affects the system only if $u_k=1$,
in which case a fixed value is added to the vertical impulse.
We wish to steer the point mass such that it passes approximately 
through a sequence of check points, as illustrated in Fig.~\ref{fig:FlappyBirdTrajectory}.

For this example, we need a slight generalization%
\footnote{This generalization is effortlessly handled by IKIE.} 
of (\ref{eqn:StateSpaceEvolution})--(\ref{eqn:trajectoryCost}) as follows.
The state $x_k \in\Reals^2$ (comprising the vertical position and the vertical speed) 
evolves according to
\begin{IEEEeqnarray}{rCl}
x_k & = & \bma 1 & T \\ 0 & 1 \ema x_{k-1}
          + \bma 0 \\ 1/m \ema u_k + \bma 0 \\ -Tg \ema,
        \IEEEeqnarraynumspace
\end{IEEEeqnarray}
and we wish the vertical position 
$y_k = \bma 1 & 0 \ema x_k$
to minimize
\begin{equation} \label{eqn:CheckpointsCost}
\sum_{k=1}^K w_k (y_k - \breve y_k)^2,
\end{equation}
where $w_k=1$ if $\breve y_k$ is a checkpoint and $w_k=0$ otherwise.

The numerical results in Fig.~\ref{fig:FlappyBirdTrajectory}
are obtained with $m=0.5$, $T=0.1$, $g=0.25$, 
$a=0$, $b=1$, $K=250$, and $s^2 = 0.1$.

\section{Conclusion}
\label{sec:Conclusion}
We have proposed a new method for controlling a linear system
with binary input, which can also be used for digital-to-analog conversion.
The key idea is a new binary-enforcing prior with a NUV representation,
which turns the actual computations into iterations of Kalman-type 
forward-backward recursions.
The computational complexity of the proposed method 
is linear in the planning horizon,
with compares favorably with existing ``optimal'' methods.

The proposed prior and method can be extended both to $M>2$ levels 
and to sparse level switching 
in the control signal, as will be detailed elsewhere.

The suitability of the proposed prior for other applications remains to be investigated.

\balance
\bibliographystyle{IEEEtran}
\bibliography{paper}

\end{document}